\documentclass[aps,10pt,nofootinbib,groupedaddress,preprintnumbers,superscriptaddress]{revtex4}

\usepackage{amssymb, amsfonts, amsmath, bm}
\usepackage[
]{graphicx}
\usepackage{color}

\usepackage{mathrsfs}

\usepackage{enumerate}
\usepackage{slashbox,ascmac}
\usepackage{ulem}

\usepackage[
]{hyperref}
\hypersetup{%
 setpagesize=false,
 bookmarksnumbered=true,%
 bookmarksopen=true,%
 colorlinks=true,%
 linkcolor=blue,%
 citecolor=red}

\begin{document}
\title{
{\Large
Effect of a second compact object on stable circular orbits
}}
\author{{\large Keisuke Nakashi}}
\email{nakashi@rikkyo.ac.jp}
\affiliation{Department of Physics, Rikkyo University, Toshima, Tokyo 171-8501, Japan}
\author{{\large Takahisa Igata}}
\email{igata@rikkyo.ac.jp}
\affiliation{Department of Physics, Rikkyo University, Toshima, Tokyo 171-8501, Japan}
\date{\today}
\preprint{RUP-19-24}

\begin{abstract}
We investigate how stable circular orbits around a main compact object appear depending on the presence of a second one by using the Majumudar--Papapetrou dihole spacetime, which consists of the two extremal Reissner--Nordstr\" om black holes with different masses.
While the parameter range of the separation of the two objects is divided due to the appearance of stable circular orbits, this division depends on its mass ratio. We show that the mass ratio range separates into four parts, and we find three critical values as the boundaries.  
\end{abstract}

\maketitle

\section{Introduction}
\label{sec:1}

Recent progress in the observation of gravitational waves supports the existence of binary black hole systems. The LIGO Scientific and Virgo collaborations have already detected gravitational waves ten times from binary black hole mergers and once from a binary neutron star merger so far~\cite{Abbott:2016blz, TheLIGOScientific:2016wfe, GBM:2017lvd,LIGOScientific:2018mvr}. Furthermore, since they have started the third observation run, the number of detections will increase in the future. These results imply that binary black hole systems are a quite common phenomenon in our Universe. 

A pure binary black hole system is a highly idealized model, around which a third object or matter distribution usually exists in realistic situations. Therefore, one of the next issues is clarifying perturbative interactions with a third body around them. As traditional problems in Newtonian gravity, there are Poincar\' e's three-body problem and the Kozai mechanism. In recent years, some problems related to these topics have been considered in the framework of the relativistic three body problem~\cite{Yamada:2010cz,Yamada:2016cnt}, in the context of resonance in a compound extreme mass ratio inspiral/massive black hole binary~\cite{Seto:2012ig}, and in gravitational wave emission induced by a third body~\cite{Suzuki:2019wvg,Wen:2002km,Seto:2013wwa,Meiron:2016ipr}.
If a third body itself is the target of observation, 
we can view it as a test body in a fixed background. 
As a traditional problem in Newtonian gravity, there is Euler's three-body problem, test particle motion in two fixed centers. The corresponding relativistic system is the main topic of this paper.

The study of test body motion is significant for the predictions of astrophysical phenomena around a gravitational system such as a binary. In particular, the circular orbit of a test body plays some essential roles in both theory and observations. In the black hole spacetime, for instance, the bending of light due to strong gravity makes a photon orbit circular near the horizon. If the circular photon orbit is unstable, it relates to the formation of the black hole shadow. On the other hand, the sequence of stable circular massive particle orbits is relevant to accretion disks and a binary system. The innermost stable circular orbit (ISCO) radius is a distinctive one because it is identified as the inner edge of a standard accretion disk model 
and a compact binary switches the stage of the evolution from the inspiraling phase to the merging phase there~\cite{Clark:1977,Kidder:1993zz}.

Actual binary black hole systems exist as highly dynamical systems so that one needs to use the numerical method to analyze the phenomena around such systems; for example, the study of the shadow of a binary black hole system requires a fully nonlinear analysis of the numerical relativity~\cite{Bohn:2014xxa}. 
On the other hand, it is also significant to use an analytical method for a qualitative understanding. 
To this end, we often employ some axisymmetric and stationary (or static) dihole spacetime as a toy model. 
There are some exact dihole spacetime solutions of the Einstein equation (or the Einstein--Maxwell equation) such as the Weyl spacetime~\cite{Weyl:1917gp}, the Majumdar--Papapetrou spacetime~\cite{Majumdar:1947eu,Papaetrou:1947ib,Hartle:1972ya}, the double-Kerr spacetime~\cite{Kramer:1980}, etc. 
We can extract the specific features of phenomena around a binary black hole system by using these dihole spacetimes. 
Indeed, the eyebrows structure of the binary black hole shadow is reproduced in the (quasi)static dihole spacetime~\cite{Nitta:2011in,Patil:2016oav,Assumpcao:2018bka,Cunha:2018cof,Shipley:2016omi}.

The aim of the present paper is to reveal how the marginally stable circular orbit (MSCO) or ISCO of the dihole spacetime varies compared to those of the single black hole spacetime. 
To achieve this, we adopt the Majumdar--Papapetrou (MP) dihole spacetime, which contains two extremal Reissner--Nordstr\"om black holes. 
The circular orbit and its stabilities in the equal mass MP dihole spacetime have been investigated~\cite{Nakashi:2019mvs, Ono:2016lql, Wunsch:2013st}. 
In our previous paper~\cite{Nakashi:2019mvs}, we clarified the dependence of the positions of MSCOs and ISCOs on the separation parameter in the equal mass MP dihole spacetime. 
We found that the range of the dihole separation is divided into five ranges and 
obtained the four critical values as the boundaries. 
In this paper, 
we investigate the sequence of the stable circular orbit in the different mass MP dihole spacetime,
which consists of the two different mass extremal Reissner--Nordstr\" om black holes. 
Once we fix the mass scale of one of the two black holes, the system depends on two parameters: the separation and the mass ratio. 
As the result of our analysis, 
we divide the mass ratio parameter range into four ranges 
and obtain three critical values as the boundaries.

This paper is organized as follows. 
In the following section, 
we introduce the MP dihole spacetime with different masses 
and derive conditions for circular particle orbits on the background. 
Furthermore, we clarify the stability conditions of these orbits 
in terms of the Hessian of a 2D potential function. 
In Sec.~\ref{sec:3}, while changing 
the mass ratio of the dihole, 
we analyze 
the dihole separation dependence of the positions of stable circular orbits. 
Due to some qualitative differences of 
sequences of stable circular orbits, 
we classify the range of dihole mass ratios into four 
parts and determine three critical values of the mass ratio 
as the boundaries of the range. 
Section~\ref{sec:4} is devoted to a summary and discussions.
Throughout this paper, we use units in which $G=1$ and $c=1$.

\section{Conditions for stable circular orbits in the Majumdar--Papapetrou dihole spacetime}
\label{sec:2}
The metric and the gauge field of the MP dihole spacetime 
in isotropic coordinates 
are given by
\begin{align}
&g_{\mu\nu}\:\!\mathrm{d}x^\mu\:\!\mathrm{d}x^\nu
=-\frac{\mathrm{d}t^2}{U^2}+U^2 (\mathrm{d}\rho^2+\rho^2\:\!\mathrm{d}\phi^2+\mathrm{d}z^2),
\\
&A_\mu\:\!\mathrm{d}x^\mu=U^{-1}\mathrm{d}t,
\\
&U(\rho, z)=1+\frac{M_+}{\sqrt{\rho^2+(z-a)^2}}+\frac{M_-}{\sqrt{\rho^2+(z+a)^2}},
\end{align}
where $M_\pm$ are each black hole mass located at $z=\pm a$~($a\geq0$). 
We introduce a mass ratio parameter 
\begin{align}
\nu:=\frac{M_-}{M_+}.
\end{align}
Without loss of generality, 
we assume that the black hole with mass $M_+$ is larger than that with mass $M_-$, i.e., 
\begin{align}
\label{eq:nurange}
0\leq \nu \leq 1.
\end{align}
We use units in which $M_+=1$ in what follows.

The Lagrangian of a particle freely falling in the MP dihole spacetime is given by
\begin{align}
\mathscr{L}=\frac{1}{2}\left[\:\!
-\frac{\dot{t}^2}{U^2}+U^2 (\dot{\rho}^2+\rho^2\dot{\phi}^2+\dot{z}^2)
\:\!\right],
\end{align}
where the dot denotes derivative with respect to an affine parameter.
Since the coordinates $t$ and $\phi$ are cyclic, 
the canonical momenta conjugate to them are constants of motion:
\begin{align}
\label{eq:EL}
E=\frac{\dot{t}}{U^2}, \quad 
L=\rho^2 U^2 \dot{\phi},
\end{align}
which are energy and angular momentum, respectively. 
We normalize the 4-velocity $\dot{x}^\mu$ so that 
$g_{\mu\nu}\dot{x}^\mu \dot{x}^\nu=-\kappa$, 
where $\kappa=1$ for a massive particle and $\kappa=0$ for a massless particle. 
Rewriting the normalization condition in terms of $E$ and $L$, we have
\begin{align}
\label{eq:energyeq}
&\dot{\rho}^2+\dot{z}^2+V=E^2,
\\
\label{eq:V}
&V(\rho, z)=\frac{L^2}{\rho^2 U^4}+\frac{\kappa}{U^2}. 
\end{align}
We can view Eq.~\eqref{eq:energyeq} as an energy equation and 
$V$ as a 2D effective potential of particle motion in the $\rho$-$z$ plane. 
In terms of $V$, the equations of motion are written as 
\begin{align}
\label{eq:eomrho}
&\ddot{\rho}+\frac{2\:\!U_z}{U}\dot{z}\:\!\dot{\rho}-\frac{2\:\!U_\rho}{U}\dot{z}^2+\frac{V_\rho}{2}=0,
\\
\label{eq:eomz}
&\ddot{z}+\frac{2\:\!U_\rho}{U}\dot{z}\:\!\dot{\rho}-\frac{2\:\!U_z}{U}\dot{\rho}^2+\frac{\:\!V_z}{2}=0,
\end{align}
where $V_i=\partial_i V$ and $U_i=\partial_i U$ ($i=\rho, z$).

We focus on circular orbits with constant $\rho$ and $z$. 
Then, the energy equation~\eqref{eq:energyeq} 
immediately reduces to
\begin{align}
\label{eq:Esq}
V=E^2. 
\end{align}
Hence, $V$ must be positive for circular orbits. 
In addition, we find that constant $(\rho, z)$ can be a solution to Eqs.~\eqref{eq:eomrho} and \eqref{eq:eomz}
when its position corresponds to an extremum of $V$: 
\begin{align}
\label{eq:Vrho}
&V_\rho=0,
\\
\label{eq:Vz}
&V_z=0.
\end{align}
We can rewrite the three conditions~\eqref{eq:Esq}--\eqref{eq:Vz}, respectively, as 
\begin{align}
\label{eq:E0}
E^2&=E_0^2(\rho, z):=V(\rho, z; L_0^2),
\\[1mm]
\label{eq:L0}
L^2&=L_0^2(\rho, z):=-\frac{\rho^3 U^2 U_\rho}{U+2\rho \:\!U_\rho},
\\
\label{eq:Uz=0}
U_z&=
\frac{a-z}{[\rho^2+(z-a)^2]^{3/2}}
-\frac{\nu(a+z)}{[\rho^2+(z+a)^2]^{3/2}}
=0.
\end{align}
From Eqs.~\eqref{eq:E0} and \eqref{eq:L0}, 
both values of $E_0^2$ and $L_0^2$ 
depend on positions of circular orbits
and must be positive.
The positivity of $L^2$ leads to that of $E^2$ as seen from Eq.~\eqref{eq:V}, 
so that it is sufficient to pay attention only to the positivity of $L^2$.

Now we solve Eq.~\eqref{eq:Uz=0}. 
If $\nu=0$, 
then we obtain the solution $z=a$.
If $0<\nu\leq 1$, 
we find from Eq.~\eqref{eq:Uz=0} that
the range of $z$ is bounded in 
$|z|<a$. 
Solving it for $\rho^2$ in this range, 
we obtain the root 
\begin{align}
\rho_0^2(z)=\frac{(a-z)^{2/3}(a+z)^2-\nu^{2/3}(a+z)^{2/3}(a-z)^2}{\nu^{2/3}(a+z)^{2/3}-(a-z)^{2/3}},\quad z\neq \frac{1-\nu}{1+\nu}a.
\end{align}
When $z=a (1-\nu)/(1+\nu)$ holds, then 
Eq.~\eqref{eq:Uz=0} 
leads to $z=0$, and hence $\nu=1$. Note that 
the root $\rho_0$
is real and positive in the range
\begin{align}
-a<z<-\frac{1-\sqrt{\nu}}{1+\sqrt{\nu}}\:\!a, 
\quad
\frac{1-\nu}{1+\nu} a<z<a.
\end{align}
The curve $\rho=\rho_0$ asymptotically approaches the line
\begin{align}
z=\frac{1-\nu}{1+\nu}\, a. 
\end{align}
In particular, the intersection point of the line with 
the symmetric axis $\rho=0$ 
corresponds to the center of mass of the dihole.
On the other hand, 
the curves terminate on $\rho=0$ 
at $z=\pm a$ (i.e., the horizons)  and
\begin{align}
\label{eq:endpoint}
z=-\frac{1-\sqrt{\nu}}{1+\sqrt{\nu}}\:\!a. 
\end{align}
Therefore, we can find a circular orbit at a point in the $\rho$-$z$ plane if 
it is located on the curve $\rho=\rho_0(z)$ and satisfies $E_0^2\geq 0$ and $L_0^2\geq0$.

To determine the stability of a circular orbit, 
we need further analysis. 
We consider the linear stability of circular particle motion 
in terms of the Hessian $V_{ij}$, where $V_{ij}=\partial_j\partial_i V$ ($i, j=\rho, z$). 
Let $h$ be its determinant, $h(\rho, z; L^2)=\mathrm{det} \:\!V_{ij}$, 
and $k$ be its trace, $k(\rho, z; L^2)=\mathrm{tr} \:\!V_{ij}$. 
By using these we define the region $D$ in the $\rho$-$z$ plane by
\begin{align}
D=\left\{ (\rho, z) \:\!|\:\! L_0^2>0, h_0>0, k_0>0 \right\},
\end{align}
where
\begin{align}
h_0(\rho, z)=h(\rho, z; L_0^2) |_{U_z=0},
\\
k_0(\rho, z)=k(\rho, z; L_0^2) |_{U_z=0},
\end{align}
where the restriction $U_z=0$ means to eliminate 
the terms proportional to $U_z$. 
We can find stable circular orbits on the curve 
$\rho=\rho_ 0 (z)$ included in the region $D$.

\section{Dependence of the sequence of stable circular orbits on the mass ratio}
\label{sec:3}

In this section, focusing on stable circular orbits in the MP dihole spacetime, 
we analyze the dependence of sequences of their orbits on the separation $a$ 
for various values of the mass ratio $\nu$. 

\subsection{$\nu=1$}
In the beginning, let us recall how sequences of 
stable circular orbits change as the separation $a$ varies 
in the equal unit mass 
MP dihole spacetime (i.e., $\nu=1$ and $M_+=M_-=1$)~\cite{Nakashi:2019mvs}.  For $a>a_0=1.401\cdots$, 
a sequence of stable circular orbits exists in the range $\rho\in (\sqrt{2}a, \infty)$ on the equidistant symmetric plane $z=0$ from each black hole. Furthermore, 
it bifurcates at $(\rho, z)=(\sqrt{2}a, 0)$
and extends towards each black hole. As a result, we have 
three MSCOs, two of which 
are the ISCOs. 
At $a=a_0$, the three 
MSCOs degenerate at 
$(\rho, z)=(\sqrt{2}a_0, 0)$. 
For $a_0\geq a>a_*=0.9713\cdots$, 
a single sequence of stable circular orbits appears on $z=0$
in the range $\rho\in (\sqrt{2}a, \infty)$, and 
this inner boundary corresponds to the ISCO. 
At $a=a_*$, 
the single sequence 
is marginally connected at a point where $h_0$ has a saddle point.
For $a_*\geq a>a_{\textrm{c}}=0.3849\cdots$, 
we have two sequences of stable circular orbits on $z=0$. 
This phenomenon implies the 
possibility of double accretion disk formation in this system. 
In particular, for $a_*>a> a_\infty=0.5433\cdots$, 
the outer sequence exists from infinity to an MSCO and 
the inner sequence from an MSCO to the ISCO, 
while for $a_\infty\geq a > a_{\textrm{c}}$, the outer boundary of the inner sequence is no longer a 
marginally stable circular massive 
particle orbit
but turns into a circular photon orbit; that is, 
infinitely large energy would be required for the stable circular orbit. 
At $a=a_{\textrm{c}}$, 
the inner sequence just disappears. 
For $a_{\textrm{c}}\geq a \geq 0$, 
we only have a single sequence of stable circular orbits from infinity to the ISCO on $z=0$.

In the following subsections, dividing the range of $\nu$ into four parts, 
we consider the dependence of sequences of stable circular orbits on the separation $a$ 
in each range of $\nu$.

\subsection{$1> \nu >\nu_\infty=0.7698\cdots$}
\label{sec:3-B}

\begin{figure}[t!]
\centering
\includegraphics[width=17cm,clip]{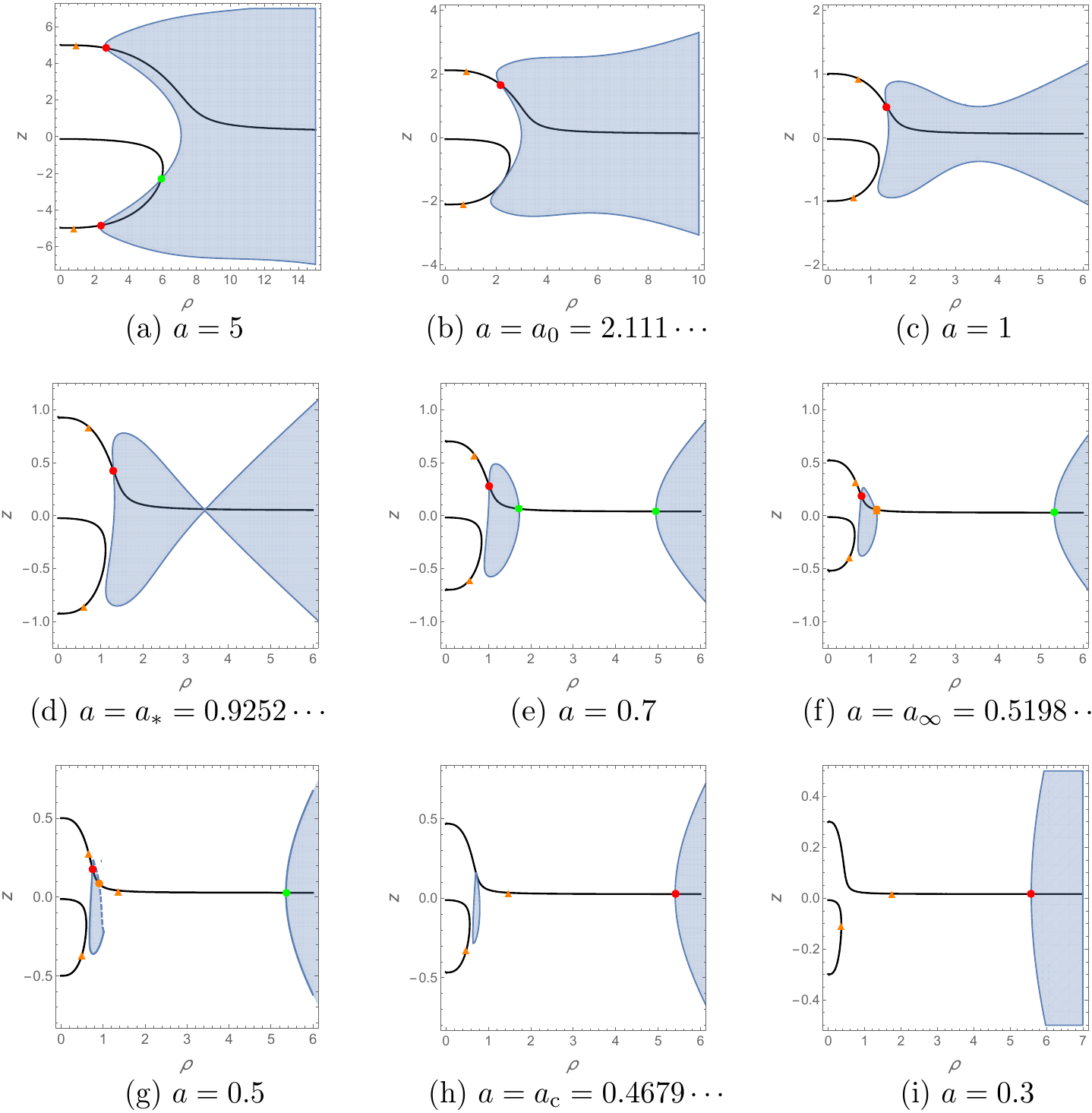}
 \caption{
Sequences of stable circular orbits in the Majumdar--Papapetrou dihole spacetime with mass ratio $\nu=0.9$. 
The black solid lines denote the curve $\rho=\rho_0$. 
The shaded regions denote the region $D$, where
$h_0>0$, $L_0^2>0$, and $k_0>0$. 
The boundaries of $D$ are shown by 
the blue solid lines on which $h_0$ vanishes
and 
the blue dashed lines on which $L_0^2$ diverges. 
The black solid lines in the shaded regions 
show the positions of stable circular orbits. 
The green dots indicate the position of 
marginally stable circular orbits, and 
the red dots indicate the position of 
the innermost stable circular orbits. 
The orange dots show the positions of 
stable circular photon orbits, and the orange triangles 
show those of unstable ones.}
 \label{fig:nu=0.9}
\end{figure}

We consider sequences of stable circular orbits 
for various values of $a$
in the MP dihole spacetime with mass ratio $\nu \simeq 1$ but $\nu\neq 1$.
We show sequences of 
stable circular orbits for several values of $a$ in the case $\nu=0.9$ in Fig.~\ref{fig:nu=0.9}. 
On the basis of these typical plots,
we discuss some qualitative properties of stable circular orbits
and critical values of $a$. 
Specific numerical values for critical values $a_0$, $a_*$, $a_\infty$, and $a_{\textrm{c}}$ in this subsection are those for $ \nu=0.9$.

For a large value of $a$, 
we have two sequences of stable circular orbits 
on both sides of the dihole~[see Fig.~\ref{fig:nu=0.9}(a)]. 
The sequence on the large black hole side 
exists from infinity to the ISCO near 
the large black hole. 
On the other hand, 
the sequence on the 
small black hole side is restricted within 
a finite region.
The inner boundary near the small black hole corresponds to the ISCO, 
and the outer boundary to an MSCO.
As the value of $a$ approaches a critical value $a_0(=2.111\cdots)$ from above, 
the MSCO
and the ISCO on the small black hole side
approach each other. 
At $a=a_0$, 
these merge into one, and then the sequence on the small black hole side
just disappears~[see Fig.~\ref{fig:nu=0.9}(b)]. 
If $a$ becomes smaller than $a_0$, 
the sequence on the small black hole side no longer exists. 
We can interpret this disappearance as 
a consequence of relativistic effects
because the corresponding sequences in Euler's three-body system,
which is governed by Newtonian gravity,
always exist for arbitrary values of $a$.

In the range $a_0\geq a>a_* (=0.9252\cdots)$, 
there exists 
a sequence of stable circular orbits only on the large black hole side, 
which appears from infinity to the ISCO~[see Fig.~\ref{fig:nu=0.9}(c)]. 
At $a=a_*$, 
the boundary of the region $D$ touches the curve $\rho=\rho_0$~[see Fig.~\ref{fig:nu=0.9}(d)]. 
This implies that 
two sequences of stable circular orbits are 
marginally connected at a point.

In the range $a_*\geq a>a_\textrm{c} (=0.4679\cdots)$, 
two sequences of stable circular orbits appear~[see Figs.~\ref{fig:nu=0.9}(d)--\ref{fig:nu=0.9}(g)].
The outer sequence exists from infinity to the outermost MSCO.
On the other hand, 
the behavior of the inner sequence divides this range of $a$ into two parts. 
For $a_*\geq a>a_\infty (=0.5198\cdots)$, 
the inner sequence exists between 
an MSCO and the ISCO. 
However, 
at $a=a_\infty$, 
the outer MSCO 
disappears because infinitely large energy 
and angular momentum would be required for a massive particle~[see Fig.~\ref{fig:nu=0.9}(f)]. 
In other words, 
a circular photon orbit appears there. 
For $a_\infty\geq a>a_{\textrm{c}}$, 
the inner sequence appears 
between the stable circular photon orbit and 
the ISCO~[see Fig.~\ref{fig:nu=0.9}(g)]. 
At $a=a_{\textrm{c}}$, 
the stable circular photon orbit and the ISCO 
merge into one, and then 
the inner sequence just disappears~[see Fig.~\ref{fig:nu=0.9}(h)]. 
In the range $a_{\textrm{c}}\geq a\geq 0$, 
there only exists a sequence of stable circular orbits, 
which appears from infinity to the ISCO~[see Fig.~\ref{fig:nu=0.9}(i)]. 

\medskip

Consequently, we divide the range of $a$ into five parts 
on the basis of typical behaviors of the sequence of stable circular orbits 
and introduce four critical values of $a$ as the boundaries of these ranges
as we have done in the case $\nu=1$. 
Note that, however, each meaning of critical values is slightly generalized from those of $\nu=1$. 
Here, let us summarize how we define the four critical values: 
\begin{itemize}
\item[(i)] $a=a_0$: The sequence of stable circular orbits on the small black hole side disappears.
\item[(ii)] $a=a_*$: The sequence of stable circular orbits on the large black hole side 
is divided into two parts.
\item[(iii)] $a=a_\infty$: 
A stable circular photon orbit appears at the outer boundary of the inner sequence of stable circular orbits on the large black hole side. 
\item[(iv)] $a=a_{\textrm{c}}$: The inner sequence of stable circular orbits 
on the large black hole side disappears. 
\end{itemize}

In the following,
according to the difference in the appearance of these critical values, 
we classify the range of the mass ratio $\nu$ into four parts. 
In each range of $\nu$, 
we discuss the behavior of the sequence of stable circular orbits depending on $a$. 
Figure~\ref{fig:ISCO}(a) shows the dependence of the radii of the MSCOs, the ISCOs, and the circular photon orbits on $a$ in the case $\nu=0.9$. 
In the range $a_\textrm{c} < a \leq a_\infty$, the radius of the ISCO (red solid line) is smaller than the one of the stable circular photon orbit (orange solid line in the middle of the three). 
In addition, 
the discontinuous transition 
of the position of the ISCO occurs 
at $a=a_\textrm{c}$. 
These phenomena are also seen in the equal mass MP dihole spacetime~\cite{Nakashi:2019mvs}.

\subsection{$\nu=\nu_\infty=0.7698\cdots$}
If we decrease the value of $\nu$ from $\nu=1$, 
then at 
\begin{align}
\nu=\nu_\infty :=\frac{4\sqrt{3}}{9}=0.7698\cdots,
\end{align}
the stable circular photon orbit no longer appears for any value of $a$. 
In other words, the critical value $a_\infty$ disappears at $\nu=\nu_\infty$. 
We can interpret that the gravity of the small black hole is not sufficiently strong to make a photon orbit circular in the region far from the large black hole even if two black holes get close each other. 
In what follows, 
we consider sequences of stable circular orbits 
in the case $\nu=\nu_\infty$.

\begin{figure}[t!]
\centering
 \includegraphics[width=5.0cm,clip]{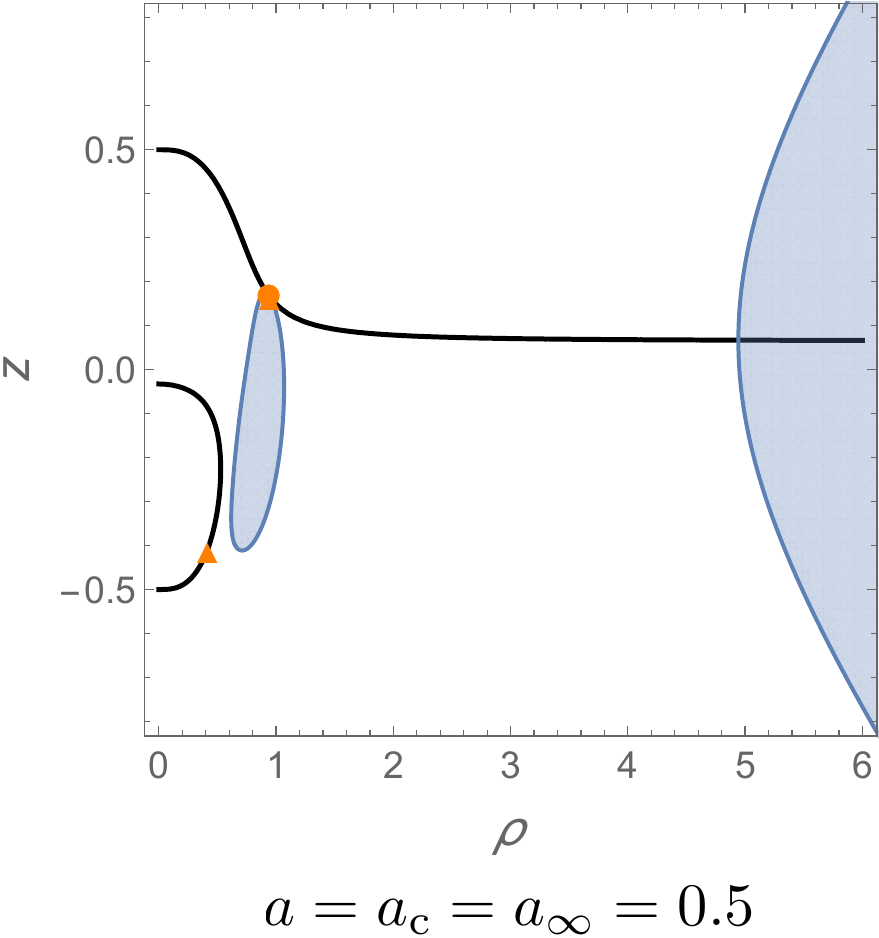}
 \caption{
Sequences of stable circular orbits in the Majumdar--Papapetrou dihole spacetime with mass ratio $\nu=\nu_\infty=0.7698\cdots$. 
The roles of each element in these plots are the same as those in Fig.~\ref{fig:nu=0.9}.
}
 \label{fig:nu=0.76}
\end{figure}

For $a>1/2$,
the behavior of sequences of stable circular orbits 
is similar as that discussed in the previous subsection. 
Indeed, we find two critical values $a_0=2.269\cdots$ and $a_*=0.8740\cdots$.
We note that, however,  
qualitative differences from the case in the previous subsection 
appear at $a=1/2$.
In the limit as 
$a \searrow 1/2$,
we find that the MSCO and the ISCO at the boundaries of 
the inner sequence merge into one at 
$(\rho,z)=(2\sqrt{2}/3, 1/6)=(0.9428\cdots, 0.1666\cdots)$~(see
Fig.~\ref{fig:nu=0.76}).
Simultaneously, 
infinitely large energy and angular momentum are required 
for a massive particle to orbit circularly here.
In other words, here is 
a stable/unstable circular photon orbit. 
These behaviors mean that $a_{\textrm{c}}$ and $a_\infty$ are degenerate at $a=1/2$, that is, $a=a_{\textrm{c}}=a_\infty=1/2$.

In the range $a<1/2$, 
there is only a single sequence of stable circular orbits that 
appears from infinity to the ISCO, 
which is the same as that discussed in the previous section. 

\subsection{$\nu_\infty>\nu>\nu_*=0.5306\cdots$}
\begin{figure}[t!]
\centering
 \includegraphics[width=17cm,clip]{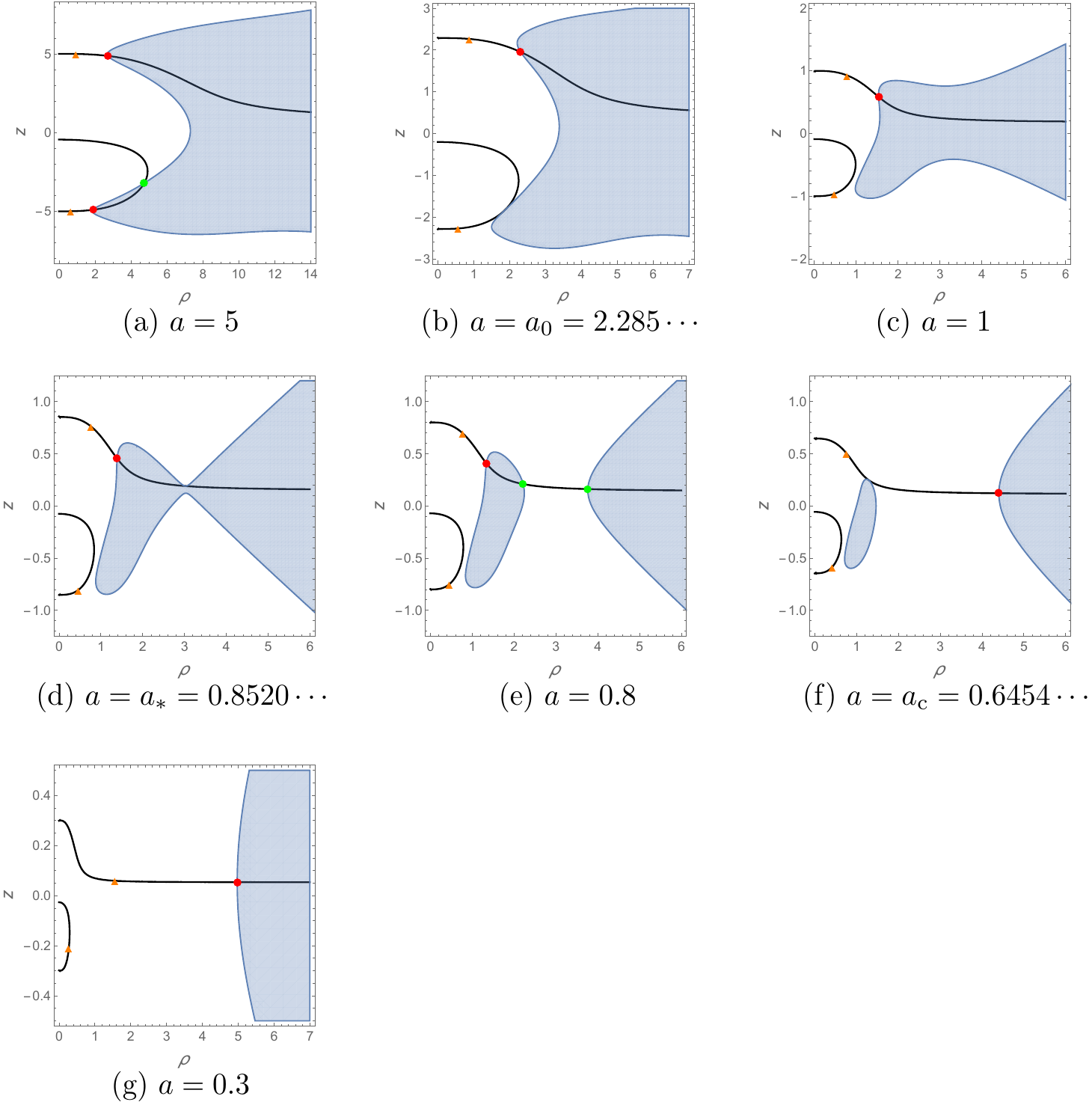}
 \caption{
Sequence of stable circular orbits in the Majumdar--Papapetrou dihole spacetime with mass ratio $\nu=0.7$. 
The roles of each element in these plots are the same as those in Fig.~\ref{fig:nu=0.9}.
}
 \label{fig:nu=0.7}
\end{figure}

We consider sequences of stable circular orbits for various values 
of $a$ in the case where $\nu_\infty>\nu>\nu_*=0.5306\cdots$. 
We can see typical sequences of stable circular orbits for 
$\nu=0.7$ in Fig.~\ref{fig:nu=0.7}. 
On the basis of these plots, we discuss 
the appearance of critical values $a_0$, $a_*$, and $a_{\textrm{c}}$ in this range. 
Specific numerical values for these critical values in this subsection 
are those for $\nu=0.7$. 

For a relatively large value of $a$, 
a sequence of stable circular orbits appears from infinity to the ISCO on 
the large black hole side, while 
a sequence appears between an MSCO and the ISCO 
on the small black hole side~[see Fig.~\ref{fig:nu=0.7}(a)]. 
When $a$ becomes smaller and smaller, at $a=a_0 (=2.285\cdots)$, the sequence on the small black hole side disappears~[see Fig.~\ref{fig:nu=0.7}(b)]. 
When $a$ becomes smaller and smaller yet, at $a=a_* (=0.8520\cdots)$, 
the sequence on the large black hole side is divided into two parts. 
In the range $a_*\geq a > a_{\textrm{c}} (=0.6454\cdots)$, 
there are two sequences, the inner and the outer. 
As a result, we find three MSCOs as the boundaries of these sequences, 
and the innermost one corresponds to the ISCO. 
At $a=a_{\textrm{c}}$, the inner sequence disappears. 
Note that the critical value $a_\infty$ no longer exists in this range of $\nu$. 
In the range $0\leq a< a_{\textrm{c}}$, 
we find a single sequence that appears from infinity to the ISCO. 
The dependence of the radii of the MSCOs, the ISCOs, and the circular photon orbits on $a$ in the case $\nu=0.7$ is shown in Fig.~\ref{fig:ISCO}(b). 
The parameter range of $a$ is divided by 
$a_0, a_*,$ and $a_\textrm{c}$
into four parts. 
The discontinuous transition of the ISCO 
on the large black hole side still occurs at $a=a_{\mathrm{c}}$. 
For any value of $a$, 
stable circular photon orbits do not exist.

\subsection{$\nu=\nu_*=0.5306\cdots$}
\begin{figure}[t!]
\centering
 \includegraphics[width=5.0cm,clip]{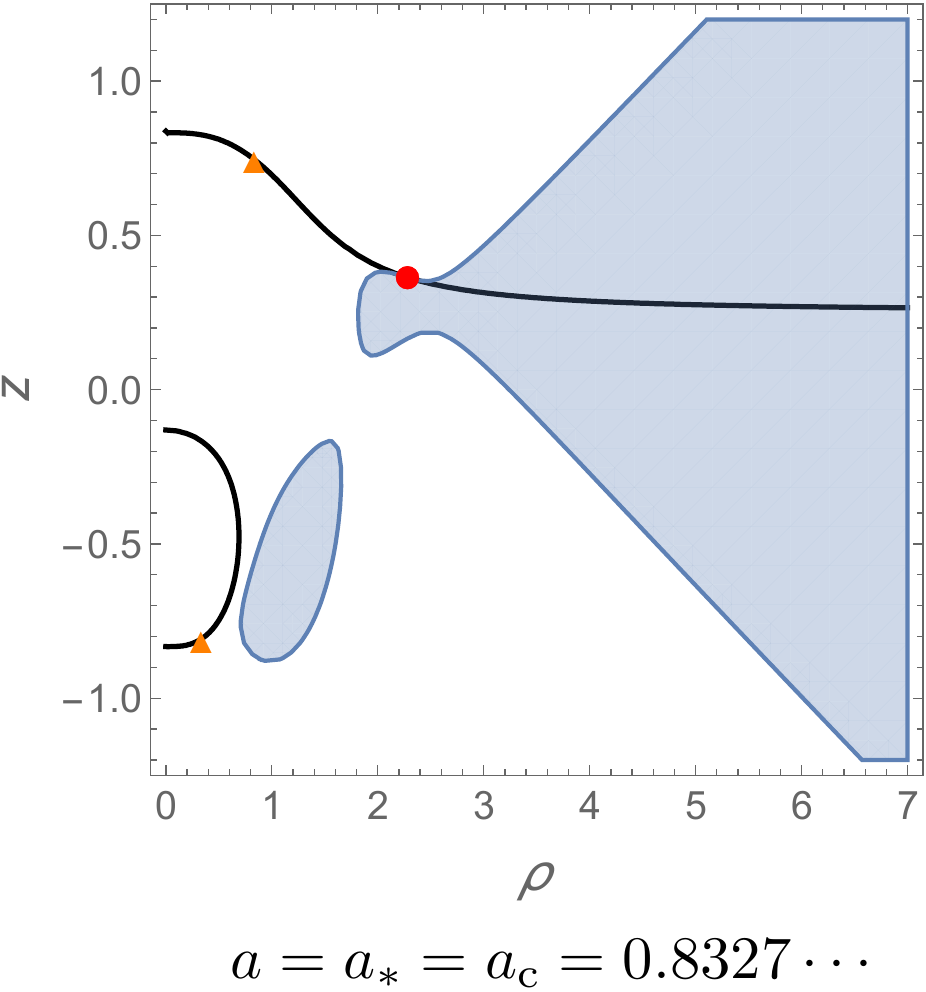}
 \caption{
Sequences of stable circular orbits in the Majumdar--Papapetrou dihole spacetime with mass ratio $\nu=\nu_*=0.5306\cdots$. 
The roles of each element in this plot are the same as those in Fig.~\ref{fig:nu=0.9}.
}
 \label{fig:nu=0.53}
\end{figure}

We focus on sequences of stable circular orbits 
for various values of $a$ in the case $\nu=\nu_*=0.5306\cdots$. 
For large $a$, we can see 
similar behavior of the sequences of stable circular orbits 
as is shown in the previous subsection. Indeed, we 
obtain the critical value $a_0=2.189\cdots$. 
We should note that 
the inner sequence of stable circular orbits appearing at $(\rho, z)=(2.279\cdots, 0.3637\cdots)$
for $a=a_*=0.8327\cdots$ 
disappears as soon as it appears~[see Fig.~\ref{fig:nu=0.53}(d)]. 
This means that the critical values $a_*$ and $a_{\textrm{c}}$ are degenerate. 
Consequently, we have no inner sequence of stable circular orbits on the large black hole side.  

\subsection{$\nu_* > \nu > \nu_0=0.0110134\cdots$}
\begin{figure}[t!]
\centering
 \includegraphics[width=17.2cm,clip]{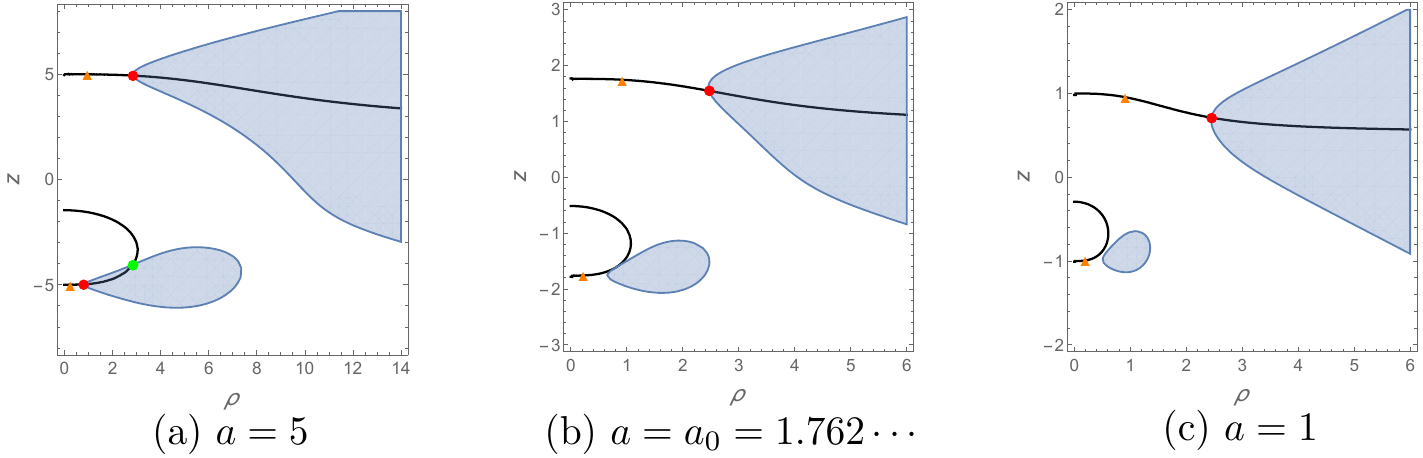}
 \caption{
Sequence of stable circular orbits in the Majumdar--Papapetrou dihole spacetime with mass ratio $\nu=0.3$. 
The roles of each element in these plots are the same as those in Fig.~\ref{fig:nu=0.9}.}
 \label{fig:nu=0.3}
\end{figure}

Let us consider sequences of stable circular orbits for various values of $a$ 
in the case where $0< \nu < \nu_*$. 
Observing typical sequences for $\nu=0.3$ in Fig.~\ref{fig:nu=0.3}, 
we discuss the appearance of the critical value $a_0$ in this range. 
The specific numerical value of $a_0$ in this subsection is that for $\nu=0.3$. 

For a large value of $a$, we find two sequences of stable circular orbits on both sides of 
the dihole~[see Fig.~\ref{fig:nu=0.3}(a)]. 
On the large black hole side, the sequence appears from infinity to the ISCO. 
On the small black hole side, 
the sequence appears from the outer MSCO to 
the ISCO. 
If $a$ becomes smaller and reaches $a=a_0 (=1.762\cdots)$, 
the sequence on the small black hole side disappears. 
Therefore, there still exists the critical value $a_0$~[see Fig.~\ref{fig:nu=0.3}(b)]. 
In the range $a<a_0$, however, 
any qualitative change of the sequence of stable circular orbits occurs
on the large black hole side. 
The dependence of the radii of the MSCOs, the ISCOs, and the circular photon orbits on $a$ in the case $\nu=0.3$ is shown in Fig.~\ref{fig:ISCO}(c).
The parameter range of $a$ is divided by $a_0$ into two parts. 
The continuous transition of the ISCO on the large black hole side no longer occurs 
because 
there are no separated sequences of stable circular orbits on the large black hole side.

\subsection{$0\leq \nu \leq \nu_0$}
\begin{figure}[t!]
\centering
 \includegraphics[width=18cm,clip]{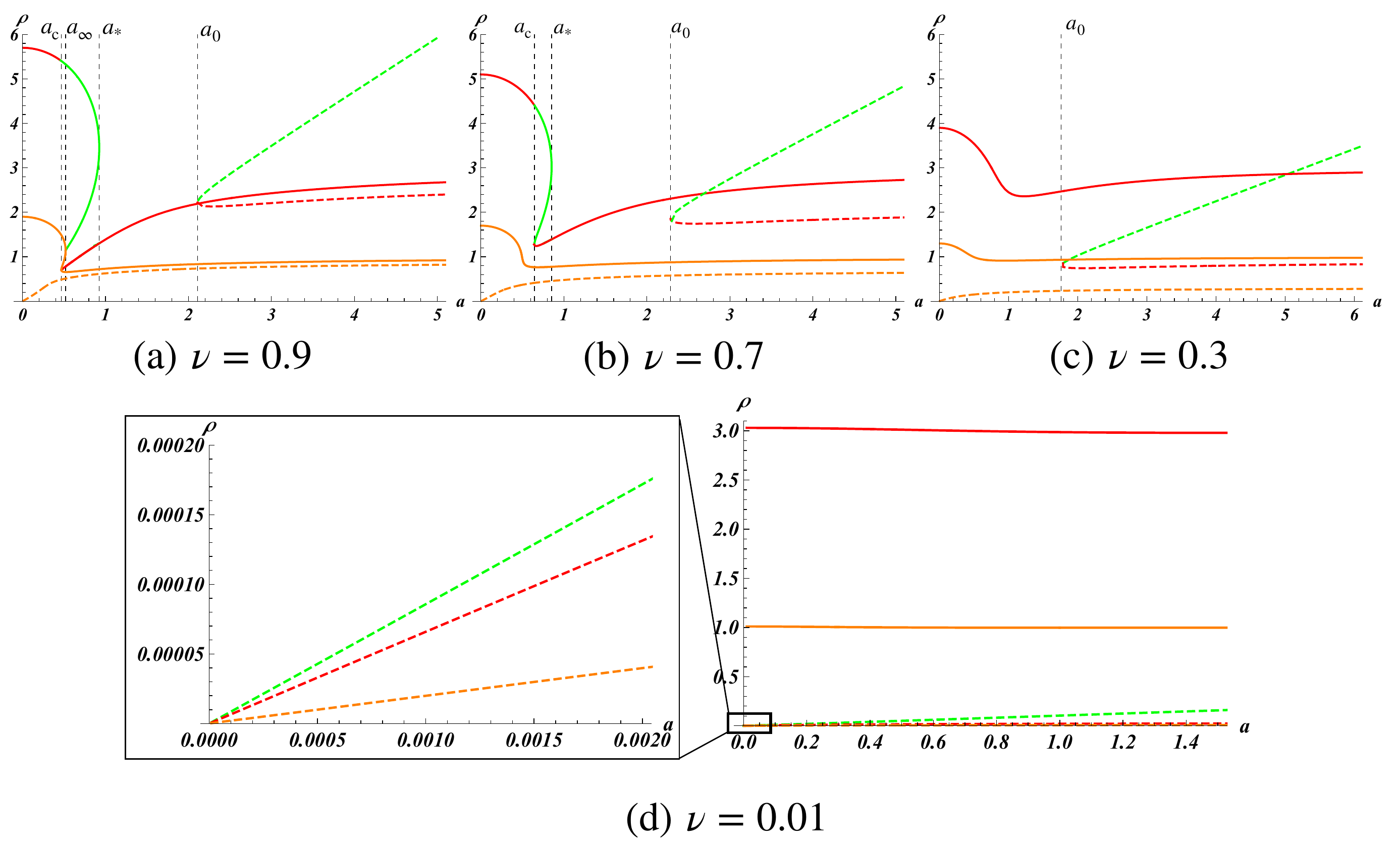}
 \caption{
Dependence of the radii of MSCOs 
 and circular photon orbits on the separation parameter $a$ 
 in the MP
 dihole spacetime with 
 mass ratio: (a) $\nu=0.9$, (b) $\nu=0.7$, (c) $\nu=0.3$, and (d) $\nu=0.01$. 
 The green and red solid lines show the radii of MSCOs 
 on the large black hole side, 
 and the green and red dashed lines show those 
 on the small black hole side. 
 In particular, the red lines indicate 
 each ISCO. 
 The orange solid lines show 
 the radii of circular photon orbits on the large black hole side, 
 and the orange dashed lines show 
 those on the small black hole side.
 The dashed green and red lines
 merge at $a=a_0$ and then the sequence of the stable circular orbits 
 on the small black hole side disappears.
In cases~(a) and (b), the green solid lines 
 emerge at $a=a_*$, and the outer exist in the range $a_{\textrm{c}}<a<a_*$.
 The inner in case~(a) exists 
 in the range $a_{\infty}<a<a_*$ while 
 the one in case~(b) exists in the range 
 $a_{\textrm{c}}<a<a_*$. 
In case~(a), 
 the stable circular photon orbits appear 
 on the large black hole side in
 $a_\mathrm{c} < a < a_\infty$ 
 whereas they do not in the other cases.
In case (d),  
the sequence on the small black hole side always exists because 
there is no critical value of $a$.}
 \label{fig:ISCO}
\end{figure}

We mention the sequence of stable circular orbits 
in the range $0\leq \nu \leq \nu_0$. 
When the value of $\nu$ reaches $\nu_0$ from above, 
the critical value $a_0$ is equal to zero. 
This means that the sequence on the small black hole side 
does not vanish unless the two black holes coalesce into one. 
If we make the value of $\nu$ smaller than $\nu_0$, 
any critical values of $a$ do not appear. 
According to Fig.~\ref{fig:ISCO}(d), where we set $\nu=0.01$, 
for $a>0$, 
both the large and the small black holes have 
the sequence of the stable circular orbits; i.e.,
two MSCOs---one of these is also the ISCO---always appear on the small black hole.

\begin{figure}[t!]
\centering
 \includegraphics[width=11cm,clip]{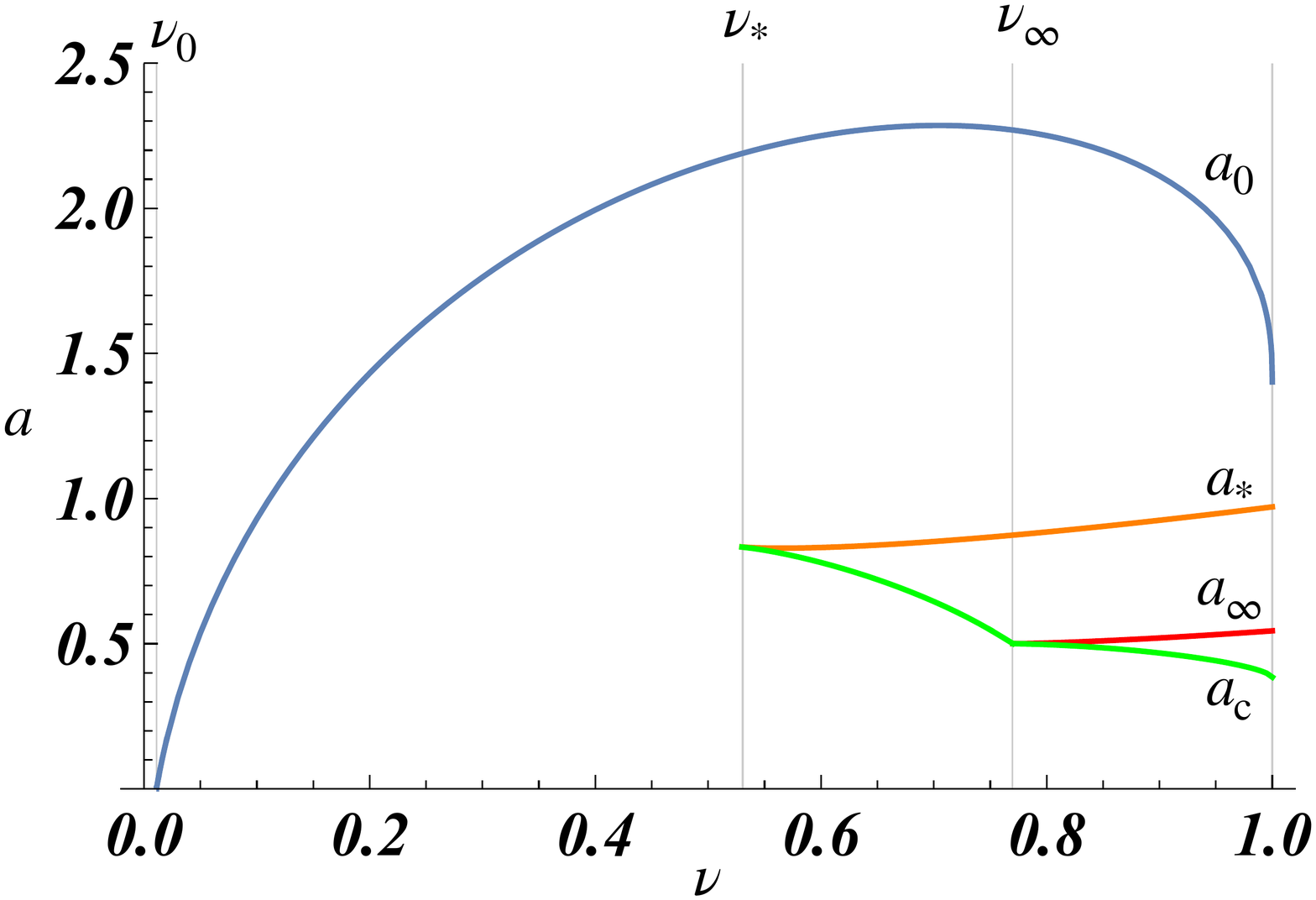}
 \caption{
 Dependence of critical values of the separation $a$ on the mass ratio $\nu$. The blue, orange, red, and green lines show $a=a_0$, $a_*$, $a_{\infty}$, and $a_{\textrm{c}}$, respectively. The parameter range of $\nu$ is divided into four parts, and the appearance of the critical values changes drastically at the boundaries, $\nu=\nu_\infty$, $\nu _*$, and $\nu_0$. At $\nu = \nu_ \infty$, 
 the critical values $a_\infty $ and $a_ \mathrm{c}$ 
 degenerate each other. At $\nu=\nu_*$, the critical values $a_*$ and $a_\mathrm{c}$ 
 coincide with each other. At $\nu=\nu_0$, the critical value $a_0$ disappears.}
 \label{fig:nu-a}
\end{figure}

\section{Summary and discussions}
\label{sec:4}

We have investigated the sequence of stable circular orbits 
around the symmetric axis
in the Majumdar--Papapetrou (MP) dihole spacetime 
with different masses.
Once we fix the mass of the large black hole to 1, 
the MP dihole spacetime is characterized 
by two parameters: 
the separation $a$ and the mass ratio $\nu$.

When $\nu \simeq 1$ but $\nu \neq 1$, 
the sequence of the stable circular orbits changes 
as $a$ varies 
in common with 
the case of $\nu=1$,
but we have generalized the definitions of the critical values of $a$ to be valid for the case of the different mass MP dihole from those in the equal mass MP dihole spacetime~\cite{Nakashi:2019mvs}. 
When the value of $a$ is relatively 
large, the sequence of 
stable circular orbits on 
the large black hole side 
exists from infinity to the ISCO 
while that on the small black hole side is restricted to a finite range. 
At $a=a_0$, 
the sequence on the small black hole side disappears. 
This phenomenon occurs due to the relativistic effect of the appearance of the ISCOs; 
that is, since the radius 
of the outer MSCO 
on the small black hole side 
decreases faster than 
the one of the ISCO as $a$ decreases, 
the positions of the MSCO and the ISCO coincide with each other at $a=a_0$, 
and then the sequence on the small black hole side disappears [see the green and red dashed lines in Figs.~\ref{fig:ISCO}(a)--\ref{fig:ISCO}(c)]. 
For $a<a_0$, 
the sequence of stable circular orbits 
appears only on 
the large black hole side. 
When $a=a_*$, 
the sequence on the large black hole side is marginally connected at a point. 
In the range $a_\mathrm{c}< a< a_*$, 
two sequences of the stable circular orbits appear 
on the large black hole side. 
The outer boundary of the inner sequence is an MSCO in $a_\infty <a < a_*$ while a stable circular photon orbit in $a_\mathrm{c} <a \leq a_\infty$. 
Finally, 
for $0\leq a \leq a_\mathrm{c}$, 
since the inner sequence vanishes, 
we have a single connected sequence from 
infinity to the ISCO 
on the large black hole side.

We have also revealed the dependence 
of the sequence of stable circular orbits 
on $\nu$. 
Figure~\ref{fig:nu-a} shows the relation between $\nu$ and the critical values of $a$. 
For $\nu > \nu_\infty = 0.7698\cdots$, 
the sequences of the stable circular orbits 
are qualitatively the same as these of the case $\nu \simeq 1$. 
At $\nu = \nu_ \infty$, 
the two critical values $a_\infty$ and $a_\mathrm{c}$ merge with each other, so that the parameter range of $a$
is divided into four parts. 
At $\nu=\nu_*=0.5306 \cdots$, 
the critical values $a_*$ and $a_\mathrm{c}$ 
coincide with each other. 
For $\nu<\nu_*$, 
the sequence on the large black hole side 
does not separate into two parts. 
The remaining critical value $a_0$ also 
disappears when $\nu=\nu_0=0.01101\cdots$. 
When we make the value of $\nu$ smaller than $\nu_0$, 
the sequences of stable circular orbits on both sides 
do not vanish until the two black 
holes merge into one [see Fig.~\ref{fig:ISCO} (d)].

The phenomena we have revealed are not caused by electric charges of the black holes, but by the presence of two black holes. 
Hence, in our Universe, we can observe such phenomena occurring around a compact object accompanied by a second compact object. 
The sequence of stable circular orbits is not on a flat plane because of the existence of a second compact object. 
Therefore, we may observe a deformed accretion disk that indicates the existence of another gravitational source.  
Furthermore, for any value of $\nu$, 
the radius of the ISCO on the large black hole side 
tends to be more inner than the case of a single black hole. 
This suggests that high energy X-rays can be detected compared to the single black hole case 
because the effective temperature of the standard disk is higher as the radius of the ISCO is smaller~\cite{Novikov:1973,Page:1974he}. 
In the ranges $\nu_*<\nu\leq 1$ and $a_{\mathrm{c}}<a<a_*$, we may observe double accretion disks. Since a relativistic effect causes the inner disk, we can use it in the testing of gravitational theories. 
The observation of 
the accretion disk of a main black hole with the second companion object
(e.g., blazar OJ287) may help us to explore the effect of the second compact object~\cite{Valtonen:2008tx}.
The presence of the stable circular photon orbit in the ranges $\nu_{\infty}<\nu\leq1$ and $a_{\mathrm{c}}< a\leq a_{\infty}$ is a characteristic property of the dihole spacetime and is associated with distinctive phenomenological features, such as the chaotic behavior of the null geodesics. We can observe qualitatively different chaotic features in the dihole shadow~\cite{Shipley:2016omi}.

Our findings indicate that the existence of the second compact object can affect 
gravitational wave emission from a test particle orbiting a main supermassive black hole
because the inspiral phase tends to be longer than the case of a single black hole because of the shift of the ISCO radius. 
In the context of the quasinormal mode, the frequency is known to
correspond to the orbital frequency of the unstable circular photon
orbit. Since our results show that the orbital frequency can be comparable to
that of a circular massive particle orbit, we can expect that the
resonant excitation of the quasinormal mode occurs~\cite{Bernard:2019nkv}.

The MP dihole spacetime we have used in the background is static, 
but a realistic binary system is a dynamical system. 
Therefore, we should take into account 
dynamical features in future work.

\begin{acknowledgments}
The authors thank T. Harada, M. Kimura, T. Kobayashi, Y. Koga, and Y. Mizuno for their fruitful discussions and useful comments. 
This work was supported by a Grant-in-Aid for Early-Career Scientists from the Japan Society for the Promotion of Science~(JSPS KAKENHI Grant No.~JP19K14715) (T.I.) and the Rikkyo University Special Fund for Research (K.N.). K.N. also thanks the Yukawa Institute for Theoretical Physics at Kyoto University, where this work was developed during the ``3rd Workshop on Gravity
and Cosmology by Young Researchers" (YITP-W-18-15).
\end{acknowledgments}

\end{document}